\documentclass[10pt,twoside]{article}
\usepackage{epsfig}
\usepackage{color}
\font\tenbf=cmbx9

\font\fourbf=cmbx12

\font\tenrm=cmr9

\font\tenit=cmti9
\font\tenitr=cmr9

\font\lrm=cmr9

\newcommand{\lle}{\mbox{$\langle$}}
\newcommand{\rle}{\mbox{$\rangle$}}

\newcommand{\bfvarepsilon}{\mbox{\boldmath$\varepsilon$}}

\newcommand{\bfsi}{\mbox{\boldmath$\sigma$}}
\newcommand{\bfep}{\mbox{\boldmath$\varepsilon$}}

\newcommand{\bfla}{\mbox{\boldmath$\lambda$}}

\newcommand{\bfzir}{\mbox{\boldmath$\bf 0$}}

\newcommand{\bfcL}{\mbox{\boldmath$\cal L$}}

\newcommand{\bfm}{\mbox{\boldmath$\bf m$}}
\newcommand{\bff}{\mbox{\boldmath$\bf f$}}
\newcommand{\bfh}{\mbox{\boldmath$\bf h$}}
\newcommand{\bfg}{\mbox{\boldmath$\bf g$}}

\newcommand{\bfn}{\mbox{\boldmath$\bf n$}}
\newcommand{\bfp}{\mbox{\boldmath$\bf p$}}

\newcommand{\bft}{\mbox{\boldmath$\bf t$}}
\newcommand{\bfs}{\mbox{\boldmath$\bf s$}}
\newcommand{\bfu}{\mbox{\boldmath$\bf u$}}

\newcommand{\bfx}{\mbox{\boldmath$\bf x$}}
\newcommand{\bfy}{\mbox{\boldmath$\bf y$}}
\newcommand{\bfz}{\mbox{\boldmath$\bf z$}}

\newcommand{\bfG}{\mbox{\boldmath$\bf G$}}

\newcommand{\bfI}{\mbox{\boldmath$\bf I$}}

\newcommand{\bfL}{\mbox{\boldmath$\bf L$}}
\newcommand{\bfN}{\mbox{\boldmath$\bf N$}}
\newcommand{\bfP}{\mbox{\boldmath$\bf P$}}

\newcommand{\bfU}{\mbox{\boldmath$\bf U$}}

\newcommand{\bfM}{\mbox{\boldmath$\bf M$}}
\newcommand{\bfT}{\mbox{\boldmath$\bf T$}}
\newcommand{\bfS}{\mbox{\boldmath$\bf S$}}

\newcommand{\bfdelta}{\mbox{\boldmath$\delta$}}

\newcommand{\bfxi}{\mbox{\boldmath$\xi$}}

\newcommand{\bfcI}{\mbox{\boldmath$\cal I$}}

\newcommand{\cV}{\mbox{$\cal V$}}

\newcommand{\bfal}{\mbox{\boldmath$\alpha$}}
\newcommand{\bfbe}{\mbox{\boldmath$\beta$}}

\newcommand{\bfGa}{\mbox{\boldmath$\Gamma$}}
\newcommand{\bfet}{\mbox{\boldmath$\eta$}}
\newcommand{\bftau}{\mbox{\boldmath$\tau$}}
\newcommand{\bfbeta}{\mbox{\boldmath$\beta$}}

\newcommand{\bfeta}{\mbox{\boldmath$\eta$}}
\newcommand{\bfom}{\mbox{\boldmath$\omega$}}

\newcommand{\BB}{\begin{equation}}
\newcommand{\EE}{\end{equation}}

\newcommand{\BBEQ}{\begin{eqnarray}}
\newcommand{\EEEQ}{\end{eqnarray}}
\begin{document}

\vspace{28pt}
\noindent {\bf Valeriy  A. Buryachenko}\footnote{\tenrm  
Department of Structural Engineering, University of Cagliari, 09124 Cagliari, Italy;
E-mail: Buryach@aol.com}

\vspace{15pt}

\noindent{\fourbf On
the thermo-elastostatics of heterogeneous materials}

\vspace{6pt}

\noindent{\fourbf  I. General integral equation}

\vspace{60pt}


\centerline{\vrule height 0.003 in width 16.cm}
 \noindent{\baselineskip=9pt
{\tenbf Abstract} {\tenrm
    We consider a linearly thermoelastic composite medium,
which consists of a homogeneous matrix containing
 a statistically inhomogeneous  random
set of inclusions, when the concentration of the  inclusions
is a function of the coordinates (so-called {\tenit
Functionally Graded Materials}). The composite medium is subjected to essentially inhomogeneous loading by the fields of the stresses, temperature and body forces
(e.g. for a centrifugal load). The general integral equations
connecting the stress and strain fields in the point being considered and the surrounding points are obtained for the random fields of inclusions. The method is based on a centering procedure of subtraction from both sides of a known initial integral equation their statistical averages obtained without any auxiliary assumptions such as, e.g., effective field hypothesis implicitly exploited in the known centering methods.
In so doing the size of a region including the inclusions acting on a separate one is finite, i.e. the locality principle takes place.
\par}}
\vspace{-5pt}

\noindent
{\lrm Keywords: A. microstructures, B. inhomogeneous material, B.
 elastic material.}

\centerline{\vrule height 0.003 in width 16.0cm}

\bigskip
\noindent {\bf 1. Introduction}
\medskip

The need for consideration  of the actual microstructure
of composite materials subjected to essentially inhomogeneous
mechanical, body force and temperature loading in
micromechanics problems is well known. Unfortunately, the
starting assumptions made in the  majority of studies,
namely that the structure of the composite media as well as the random
fields of stresses are statistically homogeneous and
therefore are invariant with respect to the translation may be invalid.

 For example, due to some production technologies,
 the inclusion concentration may be a function of the coordinates
 (see e.g. [1-3]).
The accumulation 
of damage also occurs locally in stress--concentration
regions, for example, at the tip of a macroscopic crack
(see e.g.  [4]).
Furthermore, in layered composite shells the location
of the fibers is random within the  periodic layers, and the
micromechanics equations are equations with almost periodic coefficients.
Finally, {\it Functionally Graded Materials} (FGMs) have been the subject of
intense research efforts from the mid--1980s when this term was originated in Japan
in the framework of a national project to develop heat-shielding
structural materials for the future Japanese space program.
FGM is a composite consisting of two or more phases which is fabricated
with a  spatial variation of its composition that  may improve the
structural  response (see e.g. [5, 6])
\footnote{ A popular macroscopic approach is
the modeling of FGMs as macroscopically elastic materials, in which
the material properties are graded but continuous and are described by
a local constitutive equation (see e.g.  [7, 8]),
Nevertheless,  nonlocal effects in the materials
with either statistically inhomogeneous or nonperiodic deterministic microstructure
were detected in [9]
(see also Chapter 12 in [10]).
However, the problem of estimation of effective
properties of such materials is beyond the scope of the present paper.}.
Moreover,  modern constructions from composite materials
on frequent occasions are subjected to essentially inhomogeneous
loading by fields of the stresses, temperature and body forces
(e.g. for a centrifugal load).

The final goals of micromechanical research of composites
involved in  a prediction of both the overall effective properties
 and statistical moments of stress-strain fields are based on the
approximate solution of exact initial  integral equations
connecting the random stress fields at the point being
considered and the surrounding points. This infinite system of
coupled integral equations is well-known for statistically
homogeneous composite materials subjected to homogeneous boundary
conditions (see e.g. [10-13]).
The goal of this paper  is to obtain a generalization
of these equations for the case of statistically
inhomogeneous structures of composite materials subjected to essentially inhomogeneous loading by fields of the stresses, temperature and body forces. The method is based on a centering procedure of subtraction from both sides of a known initial integral equation the statistical averages obtained without any auxiliary assumptions such as, e.g., effective field hypothesis implicitly exploited in the known centering methods. Working with statistical averages (rather than with volume averages) is convenient because statistical averaging commutes with differentiating and integrating that becomes fundamentally important for statistically inhomogeneous media.

\medskip
\noindent{\bf 2. Preliminaries}
\medskip

\noindent{\it 2.1 Basic equations}
\medskip

\setcounter{equation}{0}
\renewcommand{\theequation}{2.\arabic{equation}}
Let a linear   elastic body occupy an open simply connected bounded domain  $w\subset R^d$
 with a smooth boundary $\Gamma$ and with an indicator
function $W$ and space dimensionality $d$
 ($d=2$ and $d=3$ for 2-$D$ and 3-$D$ problems, respectively).
The domain $w$ contains  a homogeneous matrix $v^{(0)}$ and
a statistically inhomogeneous set $X=(v_i)$ of inclusions $v_i$
with indicator functions $V_i$ and bounded by the
closed smooth  surfaces $\Gamma_i$ $(i=1,2,\ldots)$.
It is assumed that the inclusions can be grouped into
    components (phases) $v^{(q)} \quad (q=1,2,\ldots,N)  $ with
 identical  mechanical and geometrical properties (such as the
 shape, size,
orientation, and microstructure of inclusions).
For the sake of definiteness, in the 2-$D$ case we will consider a
plane-strain problem.  At first no restrictions are imposed on the
  elastic symmetry of the phases or on the geometry of the inclusions.
\footnote{It is known that for 2-$D$ problems the plane-strain state
is only possible for material symmetry no lower than orthotropic
(see e.g. [14])
that will be assumed hereafter in 2-$D$ case.}
The local strain tensor
$\bfvarepsilon$  is related to the displacements
$\bf u$ via the linearized strain--displacement equation
$\bfvarepsilon ={1\over 2}[\nabla \otimes {\bf u}+(\nabla \otimes
{\bf u})^{\top}].$  Here $\otimes$ denotes
tensor product,
 and $(.)^{\top}$  denotes matrix transposition. The stress tensor,
$\bfsi $, satisfies the equilibrium equation:$\nabla\cdot \bfsi =-{\bf f}$, where
the body force tensor $\bff$ can be generated, e.g.,  by either gravitational loads or a centrifugal load. Stresses and strains are related to each other
via the constitutive equations
$\bfsi ({\bf x})={\bf L(x)}\bfvarepsilon ({\bf x})+
\bfal ({\bf x})$
or $ \bfep ({\bf x}) = {\bf M(x)}{ \bfsi}({\bf x})+\bfbe({\bf x}),$ where
 ${\bf L(x)}$ and ${\bf M(x) \equiv L(x)}^{-1}$
are the known phase stiffness and  compliance
fourth-order tensors, and the common notation for contracted products
has  been employed: ${\bf L}\bfvarepsilon =L_{ijkl}\varepsilon_{kl}$.
$\bfbeta({\bf x})$ and $\bfal({\bf x})\equiv-{\bf L(x)}\bfbeta({\bf x})$
are second-order tensors of local eigenstrains and eigenstresses.
In particular, for isotropic constituents the local stiffness
tensor $\bfL(\bfx)$ is given in terms of the local bulk modulus $k(\bfx)$ and the local
shear modulus $\mu(\bfx)$, and the local eigenstrain $\bfbe(\bfx)$ is given in
 terms of the bulk component $\beta_0(\bfx)$ by the relations:
\BB
\bfL(\bfx)=(dk,2\mu)\equiv dk(\bfx)\bfN_1+2\mu(\bfx)\bfN_2,
\quad \bfbe(\bfx)=\beta_0(\bfx)
\bfdelta,
\EE
${\bf N}_1=\bfdelta\otimes\bfdelta/d, \ {\bf N}_2={\bf I-N}_1$ $(d=2\ {\rm or}\ 3$);
$\bfdelta$ and $\bfI$ are the unit second-order and fourth-order tensors,
and $\otimes$ denotes tensor product. For the fiber composites it is the plane-strain bulk modulus $k_{[2]}$
 -- instead of the 3-D bulk modulus $k_{[3]}$ -- that plays the
significant role: $k_{[2]}=k_{[3]}+\mu_{[3]}/3$,
$\mu_{[2]}=\mu_{[3]}$.
We introduce a {\it comparison body}, whose
mechanical properties  $\bfg^c$
$(\bfg=\ \bfL, \bfM,\ \bfal, \bfbeta, \bff)$
denoted by the upper index $c$ and $\bfL^c,\bfM^c$ will usually be taken
as uniform over $w$, so that the corresponding boundary value problem is easier to solve
than that for the original body.
 All tensors ${\bf g}$ $({\bf g=L, M},\bfal, \bfbeta)$
of material properties
are decomposed as ${\bf g\equiv g}^{c}+{\bf g}_1({\bf x})=\bfg^c+
\bfg_1^{(m)}(\bfx)$ at $\bfx\in v^{(m)}$.
The upper index $^{(m)}$ indicates the
components and the lower index $i$ indicates the individual
inclusions; $v^{(0)}=w\backslash v$, $ v\equiv \cup v^{(k)}\equiv\cup v_i,
\ V(\bfx)=\sum V^{(k)}=\sum V_i(\bfx)$, and $V^{(k)}(\bfx)$ and $V_i(\bfx)$ are the
indicator functions of $v^{(k)}$ and $v_i$, respectively,  equals 1 at
$\bfx\in v^{(k)}$ and 0 otherwise, $(m=0,k;\ k=1,2,\ldots,N;
\quad i=1,2,\ldots)$.

We assume that the phases are perfectly bonded, so that
the displacements and  the traction components
are continuous across the interphase boundaries,
i.e. $[[\bfsi\bfn^{int}]]={\bf 0}$ and $[[{\bf u}]]={\bf 0}$ on the interface
boundary $\Gamma^{int}$ where ${\bf n}^{int}$ is the normal vector on
$\Gamma^{int}$ and $[[(.)]]$ is a jump operator.
The traction ${\bf t}(\bfx)=
\bfsi(\bfx)\bfn(\bfx)$ acting on any plane
with the normal $\bfn(\bfx)$ through the point $\bfx$ can be represented in
terms of displacements $\bft(\bfx)=
\hat{\bft}(\bfn,\nabla){\bf u}(\bfx)+\bfal ({\bf x})\bfn$,
where $\hat t_{ik}(\bfn,\nabla)=
L_{ijkl}n_j(\bfx){\partial/ \partial x_l}$.
The boundary conditions at the interface boundary  will be considered
together with the mixed boundary conditions on $\Gamma$
with the unit outward normal ${\bf n}^{\Gamma}$
\BBEQ
{\bf u}(\bfx)&=&{\bf u}^{\Gamma}(\bfx),\ \ \ \bfx\in \Gamma_u, \\
\bfsi(\bfx){\bf n}^{\Gamma}(\bfx)&=&{\bf t}^{\Gamma}(\bfx), \ \ \ \bfx\in \Gamma_t,
\EEEQ
where $\Gamma_u$ and $\Gamma_t$ are prescribed displacement and traction
boundaries  such that $\Gamma_u\cup\Gamma_t=
\Gamma,\ \Gamma_u\cap\Gamma_t=\emptyset$.
${\bf u}^{\Gamma}(\bfx)$ and ${\bf t}^{\Gamma}(\bfx)$ are, respectively,
 prescribed
displacement on $\Gamma_u$ and traction on $\Gamma_t$;
mixed boundary conditions, such
as in  the case of elastic supports are possible.
Of special practical interest are the homogeneous boundary conditions
\BBEQ
\bfu^{\Gamma}(\bfx)&=&\bfep^{\Gamma}\bfx,\ \bfep^{\Gamma}\equiv{\rm const.},\
\bfx\in\Gamma,
\\
{\bf t}^{\Gamma}(\bfx)&=&\bfsi^{\Gamma}{\bf n}^{\Gamma}(\bfx), \ \
 \bfsi^{\Gamma}={\rm const.},\ \bfx\in\Gamma,
\EEEQ
where $\bfep^{\Gamma}(\bfx)={1\over 2}\big[\nabla\otimes\bfu^{\Gamma}
(\bfx)+(\nabla\otimes
\bfu^{\Gamma}(\bfx))^{\top}\big],\ \bfx\in \Gamma$, and $\bfep^{\Gamma}$
and $\bfsi^{\Gamma}$ are the macroscopic strain and stress tensors, i.e. the given constant symmetric tensors.
We will consider the interior problem when the body
occupies the interior domain with respect to $\Gamma$.

\medskip
\noindent {\it 2.2 Statistical description of the composite microstructure}
\medskip

It is assumed that the representative macrodomain $w$ contains a statistically
large number of realizations $\alpha$ (providing validity of the standard probability technique) of inclusions $v_i\in v^{(k)}$ of the constituent $v^{(k)}$
$(i=1,2,\ldots;\ k=1,2,\ldots,N)$. A random parameter $\alpha$ belongs to a sample space ${\cal A}$, over which a probability density $p(\alpha)$ is defined
(see, e.g., [15, 16]).
For any given $\alpha$, any random function $\bfg(\bfx,\alpha)$ (e.g., $\bfg=V,V^{(k)},\bfsi,\bfep$) is defined explicitly as one particular member, with label $\alpha$, of an ensemble realization. Then, the mean, or ensemble average  is defined by the angle brackets enclosing the quantity $\bfg$
\BB
\lle\bfg\rle(\bfx)=\int_{\cal A}\bfg(\bfx,\alpha)p(\alpha)d\alpha. 
\EE
No confusion will arise below in notation of the random quantity $\bfg(\bfx,\alpha)$
if the label $\alpha$ is removed.
One treats two material length scales (see, e.g, [17]):
the macroscopic scale $L$, characterizing the extent of $w$, and the microscopic scale $a$, related with the heterogeneities $v_i$. Moreover, one supposes that applied field varies on a characteristic length scale $\Lambda$. The limit of our interests for both the material scales and field one is
\BB
L\gg\Lambda\geq a. 
\EE
All the random quantities  under
discussion are described by statistically inhomogeneous  random  fields.
    For the alternative description of the random structure
of a composite material let us introduce
 a conditional probability density $\varphi (v_i,{\bf x}_i
\vert v_1,{\bf x}_1, \ldots,v_n,{\bf x}_n)$,
which is a probability density to find the
$i$-th inclusion  with the center ${\bf x}_i$ in the domain
$v_i$ with fixed inclusions  $v_1,\ldots,v_n$
with the centers ${\bf x}_1,\ldots ,{\bf x}_n$. The notation
 $\varphi (v_i  , {\bf x}_i\vert ;v_1,{\bf x}_1,\ldots ,v_n,{\bf x}_n)$ denotes
the case ${\bf x}_i\neq {\bf x}_1,\ldots ,{\bf x}_n$. We will consider a general case of
statistically inhomogeneous media { with the homogeneous matrix} (for example for so-called {\it Functionally Graded Materials} (FGM)),
when the conditional probability density is  not invariant
with respect to translation:  $\varphi (v_i
 , {\bf x}_i+\bfx\vert v_1,{\bf x}_1,\ldots ,v_n,{\bf x}_n)$ $\neq
 \varphi (v_i, {\bf x}_i\vert v_1,{\bf x}_1+{\bf x},\ldots ,v_n,
{\bf x}_n+{\bf x})$, i.e. the microstructure functions depend upon
their absolute positions (see e.g. [18]).
In particular, a random field is called statistically homogeneous
in a narrow sense if its multi-point  statistical moments of
any order are shift-invariant functions of spatial variables.
Of course, $\varphi(v_i, {\bf x}_i\vert ;v_1,{\bf x}_1,\ldots
,v_n,{\bf x}_n)=0$ for values of ${\bf x}_i$ lying
inside the ``excluded volumes''
$\cup v^0_{mi}$ (since inclusions cannot overlap, $m=1, \ldots ,n)$, where  $v^0_{mi}\supset v_m$ with indicator function $V^0_{mi}$
is the ``excluded volumes'' of $\bfx_i$ with respect to $v_m$
(it is usually assumed that $v^0_{mi}\equiv v^0_m$), and
 $\varphi
(v_i, {\bf x}_i\vert ;v_1,{\bf x}_1,\ldots ,v_n,{\bf x}_n)\to \varphi(v_i,
{\bf x}_i)$
as $\vert {\bf x}_i-{\bf x}_m\vert\to \infty ,
\ m=1,\ldots,n$ (since no long-range order is assumed).
$\varphi (v_i,{\bf x})$
is a number density, $n^{(k)}=n^{(k)}({\bf x})$ of component $v^{(k)}\ni v_i$
at the point ${\bf x}$  and $c^{(k)}=c^{(k)}({\bf x})$ is
the
concentration, i.e. volume fraction, of the component $v_i\in v^{(k)}$ at the point
 ${\bf x}$:
$ c^{(k)}({\bf x})=\langle V^{(k)}\rangle ({\bf x})=\overline v_in^{(k)}({\bf x}),
\ \overline v_i={\rm mes} v_i\ \ (k=1,2,\ldots,N;\ i=1,2,\ldots),\quad
c^{(0)}({\bf x})=1-\langle V\rangle ({\bf x}).$
  The notations $\langle (.)\rangle ({\bf x})$ and
 $\langle (.)\vert v_1,{\bf x}_1;\ldots ;v_n,{\bf x}_n\rangle ({\bf x})$
 will be used for the average and for the conditional average taken
for the ensemble of a statistically inhomogeneous
 field $X=(v_i)$ at the point ${\bf x}$,
on the condition that there are inclusions at
 the points ${\bf x}_1,\ldots,{\bf x}_n$ and
${\bf x}_i\neq\bfx_j$ if $i\neq j$ ($i,j=1,\ldots, n)$.
The notations $\langle (.)\vert; v_1,{\bf x}_1;\ldots;v_n,{\bf x}_n\rangle ({\bf x})$
 are used for   the case ${\bf x}\notin v_1,\ldots,v_n$.
The notation $\lle(\cdot)\rle_i(\bfx)$ at $\bfx\in v_i\subset v^{(k)}$ means the average over an ensemble realization of surrounding inclusions  (but not over the volume $v_i$ of a particular inhomogeneity, in contrast to $\lle(\cdot)\rle_{(i)}$) at the fixed $v_i$.

\bigskip
\noindent{\it 2.3 General integral equation for composites of any structure}
\medskip

In the framework of the traditional scheme, we   introduce  a homogeneous
``comparison" body with homogeneous moduli ${\bfL}^c$, and with the
inhomogeneous deterministic transformation field $\bfbe^c(\bfx)$ and
 body force $\bff^c(\bfx)$  (and with solution $\bfsi^0,\ \bfep^0,\ {\bf u}^0$ to the same boundary-value problem). For all material tensors ${\bf g}$ (${\bfL,\bfM},\ \bfbe,\ \bfal,
\ \bff$)  the notation ${\bf g}_1(\bfx)\equiv {\bf g(x) -g}^c$ is used.

Then, substituting the constitutive equation and the Cauchy equation into
the equilibrium equation, we obtain a differential
equation with respect to the displacement $\bfu$ which
can be reduced to  a symmetrized integral form after integrating by parts
(see, e.g, [19] and
Chapter 7 in Ref. [10])
\BBEQ
{\bfep}(\bf x)&=&{\bfep}^0(\bfx)
 + \int_{w} {\bf U}(\bfx-\bfy)\bftau(\bfx)
d{\bf y}
\nonumber\\
&+&\int_w \nabla\bfG(\bfx-\bfy)\bff_1(\bfy)d\bfy+\int_{\Gamma}  \nabla {\bf G}(\bfx-\bfs)\bftau(\bfs)
{\bf n(s)}d{\bf s},
\EEEQ
 where $\bftau(\bfx)\equiv{\bfL}_1({\bf y})[{\bfep}({\bf y})-\bfbe(\bfy)]
-{\bfL}^c\bfbeta_1({\bf y})$ is called the stress polarization tensor, and  the surface integration is taken over the external surface
$\Gamma$ with the outer normal $\bfn(\bfs)\bot\Gamma$
of the macrodomain $w\subset R^d$,
and the integral operator kernel ${\bf U}$
is an even homogeneous a generalized function of degree $-d$ defined by the
second derivative of the Green tensor ${\bf G}$:
$ U_{ijkl}(\bfx)=\big[\nabla_j\nabla_l G_{ik}(\bfx)\big]_{(ij)(kl)}$,
the parentheses in indices mean symmetrization,
 and
${\bf G}$
is the infinite-homogeneous-body Green's function of the
 Navier equation with an elastic modulus tensor ${\bfL}^c$
defined by
\BB
\nabla \left\lbrace{{\bfL}^c{1\over 2}
\left[{\nabla \otimes {\bf G}(\bfx)+
 (\nabla\otimes{\bf G}(\bfx))^{\top}}\right]}\right\rbrace =-\bfdelta \delta
 ({\bf x}),
\EE
and vanishing at infinity ($|\bfx|\to\infty$),
$\delta ({\bf x})$ is the Dirac delta function and $\bfdelta$ is
the unit second order tensor. The deterministic function
${\bfep}^0(\bfx)$ is the strain field which would  exist in the medium
with homogeneous properties ${\bfL}^c$ and appropriate
boundary conditions (see, e.g, [20]):
\BBEQ
\!\!\!\!\!\!\!\!{\varepsilon}_{pq}^0(\bfx)=\int_{\Gamma}
\Big[G_{i(p,q)}(\bfx-\bfs)\b{t}_{i}(\bfs)-
u_i(\bfs)L^c_{ijkl}G_{k(p,q)l}(\bfx-\bfs)
n_j(\bfs)\Big]d\bfs 
+\int_w G_{i(p,q)}(\bfx-\bfy)f_{i}^c(\bfy)d\bfy ,
\EEEQ
which conforms with the stress field $\bfsi^0(\bfx)={\bfL}^c\bfep^0(\bfx)-\bfbe^c(\bfx)$, $\b{\bft}=\bft-\bfal^c\cdot\bfn(\bfs)$.
The representation (2.10) is valid for both the general cases of the first and
second boundary value problems as well as for the mixed boundary-value problem
(see for references [10]).
In particular, for the conditions (2.2), (2.4) and
$\bfbe^c,\bff^c\equiv\bfzir$
the right-hand-side integral over the external surface in (2.10)
can be considered as a continuation
of $\bfep ^{\Gamma}(\bfx),$ $\bfx\in \Gamma_u\equiv\Gamma$ { i.e.} (2.4), into $w$
as the strain field that the boundary condition (2.2), (2.4)
 would generate in the comparison medium with homogeneous
 moduli ${\bfL}^c$. For simplicity we will consider only internal points
$\bfx\in w$ of the microinhomogeneous macrodomain $w$ at sufficient distance from the boundary
\BB
a\ll|\bfx-\bfs|,\ \forall \bfs\in \Gamma, 
\EE
when the validity of Eq. (2.10) takes
place except in some ``boundary layer" region close to
the surface $\bfs\in\Gamma$ where some boundary data $[\bfu(\bfs),\bft(\bfs)]$ (if they are not prescribed by the boundary conditions) will depend on perturbations introduced by all inhomogeneities, and, therefore $\bfep^0(\bfx)=\bfep^0(\bfx,\alpha)$.

It should be mentioned that for the constant gravitation loads
$f_i^c=\rho^c g_i$ with a constant mass density $\rho^c$ and a constant gravitation
field $g_i$, as well as for a centrifugal load
$f_i=g_{ij}x_j$
with the matrix $g_{ij}={\rm const.}$, the volume integral
in Eq. (2.10) can be transformed into a surface integral
(see e.g. [20]).
Consider next the construction of the regularization of generalized function of
the type of derivatives of homogeneous regular function we will
use a scheme proposed by
Gel'fand and Shilov [21]
according
to which the tensor $\bfU(\bfx)$ is split into two parts
\BB
\bfU(\bfx)=\bfU^s(\bfx)+\bfU^f(\bfx),
\EE
where $\bfU^s(\bfx)=\delta(\bfx)\widetilde{\bfU}^s$,
($\widetilde{\bfU}^s\equiv {\rm const}.$) is a singular
{ function} associated with some infinitely small
exclusion region and $\bfU^f(\bfx)\equiv 1/r^{-d}
\widetilde {\bfU}^f(\bfn)$, ($\bfx=r\bfn, \ r=|\bfx|$)
is a formal function. Both terms on the right-hand side of
(2.12)  depend on an exclusion region being prescribed, while their
sum, being the left-hand side of (2.12), is defined uniquely
(see, e.g., [10]).

\setcounter{equation}{0}
\renewcommand{\theequation}{3.\arabic{equation}}
\bigskip
\noindent{\bf 3. Random structure composites}
\medskip

\noindent{\it 3.1 Known general integral equations}
\medskip

To avoid much mathematical manipulations, we will consider in this subsection the case
of a pure mechanical loading $\bff,\bfbe\equiv{\bf 0}$ and
$\bfL^c\equiv \bfL^{(0)}\equiv$const. Then for a finite number of inhomogeneities totally placed in the macrodomain $w$, Eq. (2.8) is reduced to simplified equation
\BBEQ
{\bfep}(\bf x)&=&{\bfep}^0(\bfx)
 + \int_{w} {\bf U}(\bfx-\bfy)
{\bfL}_1({\bf y}){\bfep}({\bf y})d{\bf y},
\EEEQ
where $\bfep^0(\bfx)\equiv$constant for the homogeneous boundary conditions (2.4) being considered. In early micromechanical research Eq. (3.1) was also exploited for the limiting case of a statistically homogeneous field of an infinite number of inhomogeneous in the whole space $w=R^d$. This unjustified generalization leads to well known convergence difficulties because $\bfU(\bfx-\bfy)$ is homogeneous generalized function of degree $-d$ and the integral in (3.1) is only conditionally convergent that gives no meaningful results
without additional justification for the mode of integration employed.
The nature of the conditional convergence is that the value of an integral taken over an infinite domain depends on the shape of this domain in specifying the limit process.
{Lipinski} {\it et al.} [22]
reduced the system (2.8) and (2.10) (at $\bff,\ \bfbe
 \equiv{\bf 0}$) to an equation analogous to (3.1); the subsequent convergence
 difficulty
 was overcome by the use of a self-consistent approach which
 is equivalent in fact to  the termination of the constituent
 $\widetilde{\bfU}^f(\bfn)$ (2.12) leading to the known
 convergence troubles.
 Ju and Tseng [23, 24]
 also used Eq. (3.1)
 and eliminated the difficulties induced by the dependence
 of a conditionally convergent integral on the shape of
 integration domain $w$ by the use of an assumption of this
 shape (see for details Chapter 7 in [10]).
  Fassi--Fehri {\it et al.} [25]
 postulated the size of integration domain in Eq. (3.1).
 For the purely mechanical loading ($\bff,\ \bfbe\equiv {\bf 0}$)
 Eq. (2.8) formally coincides with the analogous relations obtained
 by  {Lipinski} {\it et al.} (1995)
 by the use of Green's functions for a bounded domain $w$
 when the surface integral in the right-hand side (2.8) vanishes.
 Its implementation is not trivial because the finite-body Green's function
 ${\bf G}^w$ is generally not known, and
replacing  ${\bf G}^w$ by ${\bf G}$ (3.1) if $w$ is large enough leads to well known convergence difficulties as discussed in detail by {Willis} [13].
These difficulties mentioned above and some other ones can be avoided by a few ways.

One way of modifying such a conditionally convergent integral resulted by the long-range interactions is the so-called method of normalization (or renormalization, in analogy to its use in quantum field theory) achieved by subtracting from (3.1) the conditionally convergent behavior which is asymptotically closed to $\bfU(\bfx-\bfy){\bfL}_1({\bf y}){\bfep}({\bf y})$ at $|\bfx-\bfy|\to\infty$.
The renormalization procedure consists in subtraction out of the conditionally convergent term by making use of an expression that has the identical convergence properties as the term presenting the difficulty, but which also has a limiting value that is known.
The renormalization  method was systematically developed
for rheological problems [26],
for conductivity of random suspensions [27],
and for elasticity of composites with spherical particles [28]
in terms of perturbations introduced by the dipole strengths of inhomogeneities. However, the correct choice of a renormalizing quantity is not always straightforward that initiated Willis and Acton [29] (see also [30])
to propose an alternative method for obtaining a convergent integral expressed through the Green function.
Rigorous justification of the last approach was proposed by O'Brien [31]
by applying the divergence theorem to the boundary integral in the equation analogous to Eq. (2.8)  that leads to (for $\bfbe,\bff\equiv {\bf 0}$, $\bfep^0\equiv\lle\bfep\rle$)
\BB
{\bfep}(\bfx)={\bfep}^0 + \int_{w} \bfU(\bfx-\bfy)[
\bfL_1(\bfy)\bfep(\bfy)- \lle\bfL_1\bfep\rle]d\bfy,
\EE
where the operation of ``separate" integration of slow $\bfU(\bfx-\bfy)$ and fast $\bfL_1(\bfy)\bfep(\bfy)$ variables will be analyzed with Eq. (3.9).
Comparing of non renormalized (3.1) and renormalized (3.2) equations, McCoy [19, 32]
suggested that one can formally remove the conditionally convergent term appearing in the former by simply setting them equal to zero.
 There are well known ``noncanonical regularizations" proposed by Kr\"oner [33]
(see also Kr\"oner [34, 35])
and attributed to Kanaun [36] 
(see also [37])
\BB
\int {\bf U}(\bfx-\bfy){\bf h}~d\bfy={\bf 0},\ \ \ {\rm or}\ \ \
\int {\bf \Gamma}(\bfx-\bfy){\bf h}~d\bfy={\bf L}^c{\bf h},
\EE
and
\BB
\int {\bf U}(\bfx-\bfy){\bf h}~d\bfy={\bf M}^c{\bf h},\ \ \ {\rm or}\ \ \
\int {\bf \Gamma}(\bfx-\bfy){\bf h}~d\bfy={\bf 0},
\EE
 for the first and the second boundary-value problem, respectively;
 ${\bf h}$ is an arbitrary constant symmetric second-order tensor,
 and the integral operator kernel,
 \BB
 {\bf \Gamma} (\bfx-\bfy)\equiv -{\bf L}^c
 \left[{{\bf I}\delta ({\bf x-y})+
 {\bf U}({\bf x-y}){\bf L}^c}\right],
\EE
  called the Green stress tensor (see  [35])
is defined by the second
  derivative of the Green tensor ${\bf G}$ (2.9).
 According to (3.5), each of the relations is a consequence
of the other from the same pair, either  (3.3) or  (3.4).
In light of the note by McCoy [19, 32],
the essence of the ``noncanonical regularization" (3.3) [and (3.4)]
 used in Eq. (3.1) can be considered as some sort of the renormalization method. In actuality this regularization is  an intuitive introduction of an
operation of generalized functions  ${\bf U}$ and ${\bf \Gamma}$ on a constant symmetric tensor $\bfh\equiv {\rm const.}$
(e.g. $\bfh=\lle\bfL_1\bfep\rle\equiv$const., see for comparison Gel`fand and Shilov [21]);
Buryachenko [38]
proved that the correctness of this regularization is questionable.

The renormalized quantity in the O'Brien method (3.2) is obtained directly from the macroscopic boundary term at the homogeneous boundary conditions (2.4) that simultaneously defines both the advantage with respect to the classical renormalization method (see e.g.
[29])
and the fundamental limitation of possible generalizations of the proposed approach to both the functional graded materials and inhomogeneous external loading [because $\lle\bfL_1\bfep\rle\equiv$const. in Eq. (3.2)].

The mentioned deficiency of Eq. (3.2) could be avoided by a centering method systematically
developed by Shermergor [12]
also for statistically homogeneous media. This method was generalized in Ref. [39]
and justified by Buryachenko [38]
as applied to the FGMs. Indeed, the centering method consists in subtracting from Eq. (2.8) their statistical average yielding (at $\bff,\bfbe\equiv{\bf 0}$)
\BB
\bfep(\bfx)=\lle\bfep\rle(\bfx) + \int_w \bfU(\bfx-\bfy)[
\bfL_1(\bfy)\bfep(\bfy)- \lle\bfL_1\bfep\rle(\bfy)]d\bfy.
\EE
The statistical averages $\lle{\bfep}\rle(\bfx), \lle\bfL_1\bfep\rle(\bfy)\neq$const. contained in Eq. (3.2) allows one to apply this equation for the analyses of nonlocal effects appearing in both the FGMs and a case of inhomogeneous external loading
(see [40];
and Chapter 12 in [10]).

The original purpose of the renormalizing term was only to provide the absolute convergence of the integral in Eq. (3.1) that is archived by long-range behavior of the function
$\bfU(\bfx-\bfy)\lle\bfL_1\bfep\rle$ at $|\bfx-\bfy|\to \infty$. However, the same term is exploited in a short-range domain $|\bfx-\bfy|\le 3a$ in the vicinity of the point $\bfx\in  w$. This term is defined only by the average tensor $\lle\bfL_1\bfep\rle$ and ignores any available microtopological information (possible nonellipsoidal and laminated structure of inhomogeneities, the shape and size of excluded volume, radial distribution function, orientation, etc.). We will demonstrate in the next Subsection that Eqs. (3.2) and
(3.6) can be improved by the use of a new renormalizing term unequally determined from rigorous averaging scheme and directly dependent on the microtopological information mentioned above.

\medskip
\noindent{\it 3.2 New general integral equations}
\medskip

As noted, the prospective centering method is based on subtracting from both sides of Eq. (2.8) their statistical averages. Now we will center Eq. (2.8) by the use of statistical averages presented in a general form $\lle\bfU(\bfx-\bfy)\bfg\rle(\bfy)$,
i.e. from both sides of Eq. (2.8) their statistical averages are subtracted
\BBEQ
{\bfep}({\bf x})
\!\!\!\!&=&\!\!\!\!
\langle {\bfep}\rangle ({\bf x})
+\int_w \lle\!\lle \nabla\bfG (\bfx-\bfy)
\bff_1\rle\!\rle(\bfy)d\bfy\nonumber+\bfcI^{\Gamma}_{\epsilon}
\nonumber \\
\!\!\!\!\!\!\!\!&+&\!\!\!\!
\int_{w}\lle\!\lle{\bf U}(\bfx-\bfy)\big\{
{\bf L}_1({\bf y})[\bfep(\bfy)-\bfbe(\bfy)]
-{\bf L}^c\bfbe_1({\bf y})\big\}\rle\!\rle(\bfy)
 d{\bf y}, 
\EEEQ
where one introduces a centering operation
\BB
\lle\!\lle\bfU(\bfx-\bfy)\bfg\rle\!\rle(\bfy)\equiv
\bfU(\bfx-\bfy)\bfg(\bfy)-\lle\bfU(\bfx-\bfy)\bfg\rle(\bfy), 
\EE
and in the  right-hand-side of Eq. (3.7), the integral over the external surface
$\Gamma$
\BB
\bfcI^{\Gamma}_{\epsilon} \equiv  \int_{\Gamma} \lle\!\lle \nabla {\bf G}(\bfx-\bfs)
\big\{
{\bf L}_1({\bf s})[{\bfep}({\bf s})-\bfbe({\bf s})]
-{\bf L}_1\bfbeta^c\big\}\rle\!\rle({\bf s}){\bf n(s)}d{\bf s}+\bfep^{0}(\bfx,\alpha)-\lle\bfep^0\rle(\bfx),
\EE
can be  dropped out, because  this tensor vanishes
 at sufficient distance ${\bf x}$
from the boundary $\Gamma$ (2.11). This means that if $|\bfx-{\bf s}|$
 is large enough for $\forall {\bfs}\in \Gamma$, then  at the
portion of the smooth surface
$d{\bf s}\approx|\bfx-\bfs|^{d-1}d\bfom^s$ with a small solid angle
$d\bfom^s$ the tensor $\nabla{\bf G}(\bfx-\bfs)|\bfx-\bfs|^{d-1}$
 depends only on the solid angle $\bfom^s$  variables
and slowly varies on the portion of the surface  $d{\bf s}$;
in this sense the tensor  $\nabla{\bf G}(\bfx-\bfs)$
is called a ``slow" variable of the
solid angle $\bfom^s$ while the expression in curly brackets on
the right-hand-side integral of Eq. (3.9) is a rapidly
oscillating function on $d{\bf s}$ and is called a ``fast" variable.
Therefore we can use a rigorous theory of ``separate" integration
of ``slow" and ``fast" variables, according to which (freely speaking)
the operation of surface integration may be
regarded as averaging (see for details,
 e.g., Ref. [41]  
and its applications {Shermergor} [12]).
If (as we assume) there is no
{\it long-range} order and the function $\varphi(v_j,\bfx_j|;
v_i,\bfx_i)-\varphi(v_j,\bfx_j)$ decays at infinity (as $|\bfx_i-\bfx_j|\to \infty$)
sufficiently rapidly\footnote{\tenrm
Exponential decreasing of this function was
obtained by Willis [42]   
for spherical inclusions;
Hansen and McDonald [43],  
Torquato and Lado [44]     
proposed a faster decreasing function for
aligned fibers of circular cross-section.}
 then it  leads to a degeneration of both the surface
integral (3.9) and the summand $\bfep^{0}(\bfx,\alpha)-\lle\bfep^0\rle(\bfx)$.

  In order to express  Eq. (3.7) in terms of  stresses
we  use the identities:
\BBEQ
{\bf L}_1(\bfep-\bfbe)&=&-{\bf L}^c{\bf M}_1\bfsi,\\ 
\bfep&=&[{\bf M}^c\bfsi+\bfbe^c]+[{\bf M}_1\bfsi+\bfbe_1]. 
\EEEQ
Substituting (3.10) and (3.11) into the right-hand-side
and the left-hand-side of (3.7),
respectively, and contracting with the tensor ${\bf L}^c$ gives
  the general integral equation for stresses
 \BB
\bfsi ({\bf x})\!=\!\langle \bfsi\rangle ({\bf x})+\!\!\int_{w}\!\!\big[
\lle\!\lle {\bf \Gamma} (\bfx-\bfy) \bfeta\rle\!\rle({\bf y})\!+\!
\lle\!\lle \bfL^c\nabla\bfG (\bfx-\bfy)
\bff_1\rle\!\rle(\bfy)\big]d\bfy\!+\!
\bfcI^{\Gamma}_{\sigma}\!,
\EE
   where we define
\BBEQ
\bfcI^{\Gamma}_{\sigma}&=&\int_{\Gamma}\lle\!\lle\bfL^c\bfG(\bfx-\bfs)\bfL^c
\bfeta\rle\!\rle(\bfs)\bfn(\bfs)d\bfs,
\\ 
\bfeta({\bf y})&=&{\bf M}_1({\bf y)}\bfsi({\bf y})+\bfbeta_1({\bf y}).
\EEEQ
If we assume no long-range order,
then the tensor $\bfcI^{\Gamma}_{\sigma}$ is degenerated
and can be dropped. The tensor $\bfeta$ is called the strain polarization tensor
and is simply a notational convenience. In
(3.14) ${\bf M}_1({\bf y})$ and $\bfbeta_1({\bf y})$
are the jumps of the compliance ${\bf M}^{(k)}$  and of the eigenstrain
$\bfbeta^{(k)}$ inside  the
component $v^{(k)}$   $(k=0,\ldots,N)$  with respect to the
constant tensors ${\bf M}^c$ and $\bfbe^c$, respectively.

For convenience of the forthcoming presentation we will
recast Eq. (3.12) in another form, for which we
introduce the operation
\BB
{\bfla}^1(\bfx)={\bfla}(\bfx)-\langle\bfla\rangle_0(\bfx)
\EE
for the random function
$\bfla$ (e.g. $\bfla=\bfsi,\ \bfep,\ \bfet, \ \bff $)
with statistical average in the matrix
$\langle{\bf \bfla}\rangle_0(\bfx)$. Then Eq. (3.12) can be
rewritten in the form
 \BBEQ
\bfsi ({\bf x})=\langle \bfsi\rangle ({\bf x})+\int_{w}\big[
\lle\!\lle {\bf \Gamma} (\bfx-\bfy) \bfeta^1\rle\!\rle({\bf y})+
\lle\!\lle \bfL^c\nabla\bfG (\bfx-\bfy)
\bff_1^1\rle\!\rle(\bfy)\big]d\bfy.
\EEEQ

The general integral equations (3.7), (3.12) and (3.16)
are  new and  proposed for statistically inhomogeneous
media for the general case of inhomogeneity of tensors
$\bfbe^c(\bfx),\ \bfbe^{(0)}(\bfx),\ \bff^c(\bfx),\
\bff^{(0)}(\bfx)$ for both the  general case of the
first and second boundary value problems as well as for
the mixed boundary-value problem. In some particular cases these equations are reduced to the known ones that will be demonstrated in Subsection 3.3.

All integrals in Eqs. (3.7), (3.12) and (3.16) converge absolutely for both the statistically homogeneous and inhomogeneous random fields $X$ of inhomogeneities.
 Indeed, even for the FGMs, the term $\lle\!\lle {\bf \Gamma} (\bfx-\bfy) \bfeta\rle\!\rle({\bf y})$ (3.12) is of order $O(\vert {\bf x-y}\vert^{-2d+2})$  as $\vert {\bf x-y}
\vert \to\infty$, and the  integrals in Eqs. (3.7), (3.12)
with the kernels $\bfU$ and $\bfGa$, respectively,
converge absolutely. In a similar manner, the integrals
with the body force density converge absolutely. In fact,
the kernel $\nabla\bfG(\bfx-\bfy)$ is of order $O(|\bfx-\bfy|^{-d+1})$ as
$|\bfx-\bfy|\to\infty$, and the term $\lle\!\lle \nabla\bfG (\bfx-\bfy)
\bff_1^1\rle\!\rle(\bfy)$
tends to zero
with $|\bfx-\bfy|\to\infty$ ($\bfx\in v_i,\ \bfy\in v_j$)
as $O(|\bfx-\bfy|^{-d+1})[\varphi(v_j,\bfx_j|;
v_i,\bfx_i)-\varphi(v_j,\bfx_j)]$. For
 no {\it long-range} order assumed, the function $\varphi(v_j,\bfx_j|;
v_i,\bfx_i)-\varphi(v_j,\bfx_j)$ decays at infinity sufficiently rapidly
 and guarantees an absolute convergence of the integrals involved.
Therefore, for $\bfx\in w$ considered in Eqs. (3.7), (3.12) and (3.16) and removed far enough from the boundary $\Gamma$ ($a\ll |\bfx-\bfs|,\ \forall \bfs\in \Gamma$),
the right-hand side integrals in (3.7), (3.12) and (3.16) do not depend
on the shape and  size of the domain $w$, and they can be replaced by the
integrals over the whole space $R^d$. With this assumption we hereafter
omit explicitly denoting $R^d$ as the integration domain in the equations
\BBEQ
{\bfep}({\bf x})
\!\!\!\!&=&\!\!\!\!
\langle {\bfep}\rangle ({\bf x})
+\int [\lle\!\lle{\bf U}(\bfx-\bfy)\bftau\rle\!\rle(\bfy)+
\lle\!\lle \nabla\bfG (\bfx-\bfy)
\bff_1\rle\!\rle(\bfy)]d\bfy,\\ 
\bfsi ({\bf x})\!\!\!\!&=&\!\!\!\!\langle \bfsi\rangle ({\bf x})+\!\!\int \big[
\lle\!\lle {\bf \Gamma} (\bfx-\bfy) \bfeta\rle\!\rle({\bf y})\!+\!
\lle\!\lle \bfL^c\nabla\bfG (\bfx-\bfy)
\bff_1\rle\!\rle(\bfy)\big]d\bfy,\\
\bfsi ({\bf x})\!\!\!\!&=&\!\!\!\!\langle \bfsi\rangle ({\bf x})+\int\big[
\lle\!\lle {\bf \Gamma} (\bfx-\bfy) \bfeta^1\rle\!\rle({\bf y})+
\lle\!\lle \bfL^c\nabla\bfG (\bfx-\bfy)
\bff_1^1\rle\!\rle(\bfy)\big]d\bfy
\EEEQ
corresponding to Eqs. (3.7), (3.12), and (3.16); in a similar manner the domain $w$ can be replaced by the whole space $R^d$ and omitted.
Thus, there are no  difficulties connected with the asymptotic behavior
of the generalized functions $\nabla\nabla {\bf G}$ and
${\bf \Gamma}$ decaying at  infinity as $\vert {\bf x-y}\vert^{-d}$, and
 there is no need to postulate either the shape or the  size of the
integration domain $w$  [45] 
or to resort to either regularization [33], [36] 
or  renormalization ([28], [32], 
see also Willis [13])
of integrals  which are divergent at infinity. The rigorous
mathematical analysis of correctness of these above mentioned methods
is beyond the purpose of the current article. Nevertheless,
it should be noted that the disruption of statistical homogeneity of
media often leads to additional difficulties, whose resolution
by these known mentioned approaches appears to be  questionable
(see for details Subsection 3.2.3 in Ref. [10]).

\medskip
\noindent{\it 3.3 Some particular cases}
\medskip

 The subsequent analysis of Eqs. (3.17)-(3.19) can be done for
the comparison medium with any elastic modulus ${\bf L}^c$,
which  necessarily leads to some additional assumptions for
the structure of the strain fields in the matrix (see for details
Chapter 8 in [10]).   
Equations (3.17)-(3.19) are much easier to solve when they
contain the stress-strain fields only inside the heterogeneities.
There are two fundamentally different approaches to ensuring it.

In the first one we postulate
\BB
\bfL^c\equiv\bfL^{(0)}.
\EE
 Then the integrands with the arguments $\bfy$ in Eqs. (3.17)-(3.19)   vanish at
$\bfy\in v^{(0)}$. However, it does not guarantee a protection from the necessity of estimation of stress-strain distributions in the matrix in the general cases of both the inhomogeneous inclusions and inhomogeneous boundary conditions. Fortunately, this domain of the matrix is only located in the vicinity of a representative inhomogeneity $v_q$ (see for details [10]).   

In the second one we choose $\bfL^c$ quite arbitrarily,
and analyze Eq. (3.18) [Eq. (3.17) can be considered analogously].
Equation (3.18) being exact for any $\langle\bfeta\rangle_0(\bfx)$
can be simplified with the additional assumption that the strain polarization
tensor in the matrix $\bfeta(\bfx),\ (\bfx\in v^{(0)})$
coincides with its  statistical average in the matrix
\BB
\bfeta(\bfx)\equiv \langle \bfeta\rangle_0(\bfx), \ \bfx\in v^{(0)}.
\EE
In so doing, the assumption (3.20) is more restricted in the sense that
the assumption (3.20) yields the assumption (3.21) (the opposite is not true)
and, moreover, in such a case the exact equality
$\bfeta(\bfx)\equiv\langle\bfeta\rangle_0(\bfx)\equiv {\bf 0}$,
$\bfx\in v^{(0)}$ holds.

Equations (3.17)-(3.19) contain the general representations for statistical averages such as, e.g., $\lle\!\lle\bfU(\bfx-\bfy)\bfg\rle\!\rle(\bfy)$,
$\lle\!\lle\bfGa(\bfx-\bfy)\bfg\rle\!\rle(\bfy)$ ($\bfg=\bfL_1\bfep,\bfL_1\bfbe,\bfet$).
We will consider the particular cases of these equations obtained for the different particular approximations of statistical averages $\lle\bfU(\bfx-\bfy)\bftau\rle(\bfy)$ and
$\lle\bfGa(\bfx-\bfy)\bfeta\rle(\bfy)$ in Eq. (3.7) and (3.12), respectively. Such an analysis will be performed for Eq. (3.17) [Eqs. (3.18) and (3.19) can be considered analogously]
for no body forces acting and purely mechanical loading i.e.
\BB
 \bff(\bfx)\equiv\bfzir,\ \bfbe(\bfx)\equiv{\bf 0}
\EE
The deterministic analog of the mentioned approximations can be presented in the following forms
\BBEQ
\int\bfU(\bfx-\bfy)\bfp(\bfy)V_i(\bfy)d\bfy&=&
\bar v_i\bfU(\bfx-\bfx_i)\lle\bfp\rle_{(i)}, \\ 
\int\bfU(\bfx-\bfy)\bfp(\bfy)V_i(\bfy)d\bfy&=&
\bar v_i\bfT_i^{\epsilon}(\bfx-\bfx_i)\lle\bfp\rle_{(i)}, 
\EEEQ
where $\bfp(\bfy)$ ($\bfy\in v_i$) is some deterministic function, $v_i$ is some representative fixed heterogeneity, and the tensors
\BB\bfT^{\varepsilon}_i{\bf ( x \!\! - \!\! x}_i)\!\!=\!\!
\cases {-(\overline v_i)^{-1}\bfP_i\
&{\rm for} ${\bf x} \in v_i,$\cr
(\overline v_i)^{-1}
\smallint {\bfU(\bfx-\bfy)}V_i({\bf y})d{\bf y}
&{\rm for} $\bfx \not \in v_i,$\cr},  
\EE
have analytical representations for ellipsoidal inclusions in an isotropic matrix
(see for reference [10]),
and $\bfP_i\equiv-\smallint\bfU(\bfx-\bfy)V_i(\bfy)d\bfy\equiv \bfS_i\bfM^{(0)}\equiv$const. (for $\forall \bfx\in v_i$)  is defined by the Eshelby [46]
tensor $\bfS_i$.
 Obviously that  the equalities (3.23) and (3.24) are only asymptotically fulfilled at $|\bfx-\bfx_i|\to \infty$, and it is possible to propose a formal counterexample where an error of both approximations (3.23) and (3.24) equals infinity, e.g. if
$\bfp(\bfx)\not\equiv {\bf 0}$ and
$\lle\bfp\rle_{(i)}={\bf 0}$ ($\bfx\in v_i)$.
The most popular approximation (3.23) (which is simultaneously the most crude)
was implicitley used by many authors (see for early references, e.g., [47])
including implied exploiting of Eq. (3.6) for obtaining of some sort of the centered Eq.  (3.2) (see [30]).   
A quantative analysis of results obtained by the use of the representations (3.23) and (3.24) will be performed in an accompanied paper by Buryachenko [48]).

Substitution of the random analog [e.g., when $\bfp(\bfx)$ is replaced by
$\bfg(\bfx,\alpha)=\bftau(\bfx,\alpha), \bfet(\bfx,\alpha)$]
of the
approximation (3.23) into Eq. (3.17)
for statistically homogeneous media subjected to the homogeneous boundary conditions (2.4) yields the known Eq. (3.2) with the renormalizing term obtained by O'Brian [31] at $\bfbe(\bfx)\equiv {\bf 0}$ 
 through the homogeneous boundary conditions.  Moreover, for the mentioned homogeneous boundary conditions and statistically homogeneous media
($n^{(q)}(\bfx)=n^{(q)}\equiv $const., $q=1,\ldots,N$), the approximations (3.23) and (3.24) leads to an identical result reducing Eq. (3.17) to  (3.2). This statement holds if we prove that contributions made to Eq. (3.17) by the renormalizing terms (3.23) and (3.24) are identical for any macrodomain $\bfx\in w$ ($\bfbe(\bfx)\equiv {\bf 0}$):
\BB
\int_w\bfU(\bfx-\bfy)d\bfy\lle \bfL_1\bfep\rle =\sum_{q=1}^N\int_w \bfT_q^{\varepsilon}(\bfx-\bfx_q)\overline{v}_qn^{(q)}(\bfx_q)\lle \bfL_1\bfep\rle_q d\bfx_q 
\EE
For justification of the equality (3.26), it should be mentioned that for uniform distribution of inclusion centers $\bfx_q$, all volume of the domain $w$ in the right-hand side of Eq. (3.26) is uniformly covered by the moving ellipsoids $v_q$. Then any point in the domain $\bfx\in w$ in the right-hand side integral (3.26) is covered by the same number $k$ of the ellipsoids $v_q$ with homogeneous strain polarization tensor $\bftau(\bfy)\equiv\lle\bfL_1\bfep\rle_q$ ($\bfy\in v_q$), and, therefore, the integral over the covered domain $w$ on the right-hand side of Eq. (3.26) is equal (within some probability factor) to
$k$ integrals over domain $w$ in the left-hand side. Therefore, in the case $\bftau(\bfy)\equiv$const. inside moving inhomogeneity $\bfy\in v_q$, both approximation (3.23) and (3.24) reduce Eq. (3.17) to the known one (3.2). However, a condition of homogeneity $\bftau(\bfy)\equiv$const. at $\bfy\in v_q$ is fulfilled only for homogeneous ellipsoidal inhomogeneities in the framework of an additional hypothesis of effective field homogeneity
according to which each inclusion is located inside a homogeneous so–called effective field
(see also [48]). 
An abandonment from effective field hypothesis  leads with necessity to inhomogeneity of the stress-strain fields inside the inhomogeneities that can tend to the different predictions of effective moduli based on Eqs. (3.2) and (3.17) even for both the statistically homogeneous media and homogeneous boundary conditions (see for details [48]).
This difference is a result of insensitivity of the renormalizing term
$\bfU(\bfx-\bfy)\lle\bfL_1\bfep\rle$ [obtained at the approximation (3.23)]
in the correct equation (3.2) to the details of heterogeneities of the stress-strain fields inside the inclusions, while a corresponding term $\lle\bfU(\bfx-\bfy)\bfL_1\bfep\rle(\bfy)$ [which is exact and obtained without approximations neither (3.23) nor (3.24)] of Eq. (3.17) explicitly depends on the mentioned field inhomogeneity.

It should be mentioned that the equality used in Eq. (3.26) (e.g., $\bfg=\bfL_1\bfep$)
\BB
\lle\bfg\rle=\sum^N_{q=1} c^{(q)}\lle\bfg\rle_q 
\EE
is only fulfilled for statistically homogeneous media subjected to the homogeneous boundary conditions; here the summation in the right-hand side is performed over the volume of the representative inclusions $v_q\in v^{(q)}$ ($q=1,\ldots,N)$.
If any of these conditions is broken then it is necessary to consider two sorts of conditional averages (see for details [10]).   
At first, the conditional statistical average in the inclusion
phase $\lle\bfg \rle^{(q)}(\bfx)\equiv\lle\bfg V\rle^{(q)}(\bfx)$ (at the condition that the point
$\bfx$ is located in the inclusion phase $\bfx\in v^{(q)}$) can be found as
$\lle\bfg V\rle^{(q)}(\bfx)=\lle V^{(q)}(\bfx)\rle^{-1}\lle\bfg V^{(q)}\rle(\bfx)$.
Usually, it is simpler to estimate the conditional averages
of these tensors in the concrete point $\bfx$ of the fixed inclusion $\bfx\in v_q$:
$\lle\bfg| v_q,\bfx_q\rle(\bfx)\equiv \lle \bfg\rle_q(\bfx)$.
Although in a general case
\BB
\lle\bfg\rle(\bfx)\equiv \sum^N_{q=1}c^{(q)} \lle\bfg\rle^{(q)}(\bfx)\not =
\sum^N_{q=1} c^{(q)} \lle\bfg| v_q,\bfx_q\rle(\bfx), 
\EE
where $v_q\in v^{(q)}$, it can be easy to establish a straightforward relation between these averages for the ellipsoidal inclusions $v_q$ with the semi-axes ${\bf a}_q=(a_q^1,\ldots,a_q^d)^{\top}$. Indeed, at first we built some auxiliary set $v_q^1(\bfx)$
 with the boundary $\partial v^1_q({\bf x})$ formed by the centers of translated ellipsoids $v_q({\bf 0})$
around the fixed point $\bfx$.
 We construct $v_q^1(\bfx)$ as a limit $v_{kq}^0\to v_q^1({\bf x})$ if a fixed ellipsoid $v_k$ is shrinking to the point $\bfx$. Then we can get a relation between the mentioned averages [$\bfx=(x_1,\ldots,x_d)^{\top}$]:
\BB
\lle  \bfg\rle^{(q)}(\bfx)=\int_{ v^1_q({\bf x},{\bf a}_q)}
n^{(q)}(\bfy)\lle \bfg|v_q,\bfy\rle(\bfx) ~d\bfy.  
\EE
 Formula (3.29) is valid for any material inhomogeneity of inclusions of any concentration in the macrodomain $w$ of any shape (if $v_q^1({\bf x})\subset w$). Obviously, the general Eq. (3.29) is reduced to Eq. (3.27) for  both the statistically homogeneous media subjected to homogeneous boundary conditions and statistically homogeneous fields $\bfg$ (e.g., $\bfg=\bfsi,\bfep$).


Thus, we have performed a qualitative competitive analysis of both the known (3.2), (3.6) and new (3.17), (3.19) general integral equations. Quantitative correlation of some estimations obtained at the bases of these equations is presented by Buryachenko [48].
Interested readers are referred to the book by Buryachenko [10]
for detailed comparison of Eqs. (3.2) and (3.6) with the related equations and approaches. It should be mentioned, that the particular cases of Eq. (3.6)
 were also widely used (either explicitly or implicitly) by other authors
 (see, e.g., [15], [16], [49-52]).
However,  Eq. (3.6) were obtained
 in the mentioned papers by a
 centering method based on subtracting from Eq. (3.1)
 [rather than from Eq. (2.8)]
 their statistical average obtained in the framework of implicit use of the asymptotic approximation (3.23) (although this approximation was not indicated).

Lastly, we will consider the field $X$ bounded in one direction
such as a laminated structure of some real FGM
(see [1], [2]).
Then the surface integral (2.8) over a ``cylindrical" surface (with the surface area
proportional to $\rho=|\bfx-{\bf s}|$) tends to zero with $|\bfx-{\bf s}|\to
\infty$ as $\rho^{-d+2}$ simply because the generalized function
$\nabla{\bf G}(\bfx-{\bf s})$ is an even homogeneous function of
order $-d+1$. Therefore, for infinite media the surface integral (2.8)
 vanishes, and Eq. (2.8) can be rewritten as
\BBEQ
{\bfep}({\bf x})=   {\bfep}^0  ({\bf x})
+\int \nabla\bfG(\bfx-\bfy)\bff_1(\bfy)d\bfy
+\int{\bf U}(\bfx-\bfy)
\{{\bf L}_1({\bf y})[{\bfep}({\bf y})-\bfbe(\bfy)]
-{\bf L}^{c}\bfbeta_1({\bf y})\}d\bfy,
\EEEQ
or, alternatively, in terms of stresses
\BB
\bfsi ({\bf x})=\bfsi^0({\bf x})
+\int\bfL^c\nabla\bfG(\bfx-\bfy)
\bff_1(\bfy)d\bfy  
+\int{\bf \Gamma}{\bf (x-y)} \bfeta({\bf y}) d{\bf y}.
\EE
In so doing the integrals from body forces in Eqs. (3.30) and
(3.31) only conditionally converge for a general case
of a bounded function $\bff_1(\bfy)$; because of this,
for the function $\bff(\bfy)$ we will assume decay at infinity
$|\bfx-\bfy|\to\infty$ no less then $O(|\bfx-\bfy|^{-\beta})$, ($\beta>0$)
guaranteeing the absolute convergence of body force integrals
in Eqs. (3.30) and (3.31).
Clearly in the considered case of   $X$ bounded in one direction,
Eqs. (3.30) and (3.31) are exact, and the right-hand-side integrals
in  (3.30) and (3.31) converge absolutely.
Eq. (3.30) was used by  Torquato [17], [53]    
for the particular case
(3.30) with homogeneous boundary conditions (2.4) and for
the inclusion field $X$ with a constant concentration of
inclusions within  an ellipsoidal domain included in the
infinite matrix. Although Eqs. (3.6), (3.17) and (3.19) are more complicated
then Eqs. (3.30) and (3.31), nevertheless they provide practical advantages
because their integrands decay at infinity faster then the integrands involved in Eq.
(3.30) and (3.31).

It should be noted that Eqs. (3.17)-(3.19) exploiting the infinite-homogeneous-body Green functions were obtained from Eqs. (3.7), (3.12), and (3.16), respectively, at sufficient distance $\bfx$ from the boundary $\Gamma$ (2.11), and, therefore, they can not be used for analysis of boundary layer effects (e.g. free edge effect). For this class of problems, the Green functions for finite domains seem more prospective (see, e.g., [15], [54] and Chapter 14 in [10]).  In such a case, the appropriate general integral equations generalizing Eqs. (3.17)-(3.19) to the finite domains can be obtained in a straightforward manner (see for details Chapter 14 in [10]). However, more detailed considerations of boundary layer effects are beyond the scope of the current study and will be analyzed in other publications.

\bigskip
\noindent {\fourbf 4. Random structure composites with long-range order}
\medskip

\setcounter{equation}{0}
\renewcommand{\theequation}{4.\arabic{equation}}
Localized Eqs. (3.17) and (3.18) were obtained in the framework of
no long-range order assumption when the integrand in curly brackets
decays at infinity $|\bfx-\bfs|\to \infty$ sufficiently rapidly.
Now we relax this assumption and for the sake of definiteness,
we will consider some conditional averages of the surface
integral in Eq. (3.18)
\BBEQ
\langle\bfcL^{\Gamma}_{\sigma}\vert v_1,{\bf x}_1;\ldots ;v_n,{\bf x}_n
\rangle ({\bf x})=
\int_{\Gamma}\Big\{\langle {\bfGa}^{\Gamma}{\bf (x-s)}\bfet^1(\bfs)
\vert;v_1,{\bf x}_1;\ldots;v_n,{\bf x}_n\rangle ({\bf s})
-\langle {\bfGa}^{\Gamma}{\bf (x-s)}\bfet^1(\bfs)\rangle ({\bf s})\Big\} d{\bf s},
\EEEQ
where $\bfx\in v_1,\ldots,v_n, \ (n=1,2,\ldots),\ \bfx\not\in \Gamma$ and
${\bfGa}^{\Gamma}{\bf (x-s)}=
-\bfL^c\nabla\bfG(\bfx-\bfs)\bfL^c$.
The asymptotic behavior
of the integrand in curly brackets in Eq. (4.1) as $|\bfx-\bfs|
\to\infty$ can be estimated by the use of representation of
the solution $\langle \bfet^1
\vert;v_1,{\bf x}_1;\ldots;v_n,{\bf x}_n\rangle ({\bf s})$
by the successive approximation method (see for details [10]).
Then
\BBEQ\langle {\bfGa}^{\Gamma}{\bf (x-s)}\bfet^1(\bfs)
\!\!\!\!\!\!\!\!&\vert&\!\!\!\!\!\!\!\!;v_1,{\bf x}_1;\ldots;v_n,{\bf x}_n\rangle ({\bf s})
-\langle {\bfGa}^{\Gamma}{\bf (x-s)}\bfet^1(\bfs)\rangle ({\bf s})
\nonumber \\
\!\!\!\!\!\!&\to&\!\!\!\!\!\langle {\bfGa}^{\Gamma}\!{\bf (x-s)}\bfet^1(\bfs)\rangle_i ({\bf s})[\varphi (v_i,{\bf x}_i
\vert v_1,{\bf x}_1, \ldots,v_n,{\bf x}_n)-\varphi (v_i,{\bf x}_i)]
\nonumber\\
\!\!\!\!&+&\!\!\!\! O( r^{-2d+1})\sum_{j=1}^n\langle \bfet^1\rangle_j (\bfx_j)
\varphi (v_i,{\bf x}_i
\vert v_1,{\bf x}_1, \ldots,v_n,{\bf x}_n),
\EEEQ
where $\bfs\in v_i$, $r=\min |\bfx_j-\bfs|,\
(j=1,\ldots,n)$ and the terms in Eq. (4.2) of order
$O(r^{-3d+1})$ and higher order terms are dropped.
The contribution of terms in  (4.2) proportional
to $O(r^{-2d+1})$ in the integral (4.1) are degenerated
at $|\bfx_j-\bfs|\to\infty$ and Eq. (4.1) can be simplified
\BBEQ
\langle\bfcL^{\Gamma}_{\sigma}\vert
v_1,{\bf x}_1;\ldots
;v_n,{\bf x}_n
\rangle ({\bf x})=
\int_{\Gamma}\lle{\bfGa}^{\Gamma}{\bf (x-s)}
 \bfet^1(\bfs)\rangle_i ({\bf s})
[\varphi (v_i,{\bf x}_i
\vert v_1,{\bf x}_1, \ldots,v_n,{\bf x}_n)
-\varphi (v_i,{\bf x}_i)]
d\bfs.
\EEEQ
If boundary conditions are applied for which
$\langle \bfsi\rangle(\bfs)$ and, therefore,
$\langle \bfet^1\rangle_j (\bfs)$ vary linearly
(or higher) with $\bfs$ and $[\varphi (v_i,{\bf x}_i
\vert v_1,{\bf x}_1, \ldots,v_n,{\bf x}_n)
-\varphi (v_i,{\bf x}_i)]$ does not decay
sufficiently rapidly as $|\bfx_j-\bfs|\to \infty$
($j=1,\ldots,n$) (long-range order) then the integral
(4.3) may be divergent. We will consider the
interesting practical case of a random structure composite described
as either a triply or a doubly periodic in the broad sense
random field $X$.

Namely, it is now assumed that the representative macrodomain $w$ contains a statistically
large number of realizations of ellipsoidal inclusions $v_i\in v^{(1)}\subset R^d \ (i=1,2,\ldots)$
with identical shape, orientation and mechanical properties.
The composite material is constructed using the
 building blocks or  cells:
$w=\cup\Omega_{\bf m},\ v_{\bf m}\subset \Omega_{\bf m}$.
We consider a composite medium with each random realization of particle centers
distributed at the nodes of some spatial lattice $\Lambda$
with the nodes $\bfm^j$ corresponding to $j=1,\ldots,n$ particles in each cell $\Omega_{\bf m}$. Suppose ${\bf e}_i$ $(i=1,\ldots,d)$ are linearly--independent vectors,
so that we can represent any node ${\bf m}\in\Lambda$ of both the triply and doubly periodic lattice as
\BB
\bfx_{\bf m}=\sum_{j=1}^n\sum_{i=1}^d m_i^j{\bf e}_i,\ \ \
\bfx_{\bf m}=\sum_{j=1}^n\Big[\sum_{i=1}^{d-1} m_i^j{\bf e}_i +f_d^j(m_d){\bf e}_d\Big],
\EE
where ${\bf m}^j=(m^j_1,\ldots,m^j_d)$ are
integer--valued coordinates of the node ${\bf m}^j$ in the  basis
${\bf e}_i$ which are equal in modulus to $| {\bf e}_i |$, and
$f_d^j(m_i)-f_d^j(m_i+1)\not \equiv {\rm const}., \ (i=1,\ldots,d).$
In the plane $f(m_d)={\rm const.}$ the composite is reinforced by
periodic arrays $\Lambda_{m_d}$ of inclusions in
the direction of the ${\bf e}_1$ axis and the ${\bf e}_{d-1}$ axis.
The type of the lattice $\Lambda_{m_d}$ is defined by the
law governing the variation in the coefficients $m_i$ ($i=1,2)$, and also
by the magnitude and orientation of the vectors ${\bf e}_i$ ($i=1,d-1)$.
In the functionally graded
direction ${\bf e}_{d}$ the inclusion spacing between adjacent arrays may vary
($f^j_d(m_d)-f^j_d(m_d+1)\not \equiv{\rm const.}).$ For a doubly--periodic
array of inclusions in a finite ply  containing $2m^l+1$
layers of  inclusions  we have $f^j(m_d)\equiv 0$ at $|m_d| > m^l$;
in more general case of
doubly periodic structures $f^j(m_d)\not\equiv 0$ at $m_d\to\pm\infty$.
To make the exposition more clear  we will assume that the basis
${\bf e}_i$ is an orthogonal one and the axes ${\bf e}_i$ ($i=1,\ldots,d$)
are directed along axes of the global Cartesian coordinate system (these
assumptions are not obligatory).

Let $\cV_{\bf x}$ be a ``moving averaging" cell with the center $\bfx$
and
 characteristic size $a_{{\cV}}=\sqrt[d]{\overline{\cV}}$, and let
 for the sake of definiteness
 $\bfxi$ be a random vector uniformly distributed on $\cV_{\bf x}$
whose value at $\bfz\in \cV_{\bf x}$ is
$\varphi_{\small \bfxi}(\bfz)=1/\overline {\cV}_{\bf x}$ and
$\varphi_{\small \bfxi}(\bfz)\equiv 0$ otherwise. Then
we can define the average of the function ${\bf g}$ with respect to
translations of the vector $\bfxi$ for each random realization of the function $\bfg$
\BB
  \langle {\bf g} \rangle _{\bf x}(\bfx-\bfy)={1\over\overline
 {\cV}_{\bf x}}\int_{\cV_{\rm \bf X}}{\bf g}(\bfz -\bfy)d\bfz,
\quad \bfx\in {\Omega} _i.
\EE
Among other things, ``moving averaging" cell $\cV_{\bf x}$
can be obtained by translation of a cell $\Omega_i$
and can vary in size and  shape  during the motion from point
to point.
 Clearly, contracting the cell $\cV_{\bf x}$ to the point $\bfx$
occurs in passing  to the limit
$  \langle {\bf g} \rangle _{\bf x}(\bfx-\bfy)\to {\bf g}(\bfx-\bfy)$.
 To make the exposition more clear we will
assume that $\cV_{\bf x}$ results from $\Omega_i$ by translation of the vector
$\bfx-\bfx^{\Omega}_i$; it can be seen, however, that this assumption is not
 mandatory.

The surface integral (4.3) can be eliminated in the equation related with (3.12)
by ``centrification" achieved  by subtracting from both sides of Eq. (3.12)
their averages over the moving averaging cell $\cV_{\bf x}$ (4.5).
In so doing the average operator (4.5) introduced for a
deterministic function $\bfg(\bfy)$ should be recast for random
function $\bfg(\bfy)$ by the use of a previous estimation
of a statistical average $\langle\bfg\rangle(\bfz-\bfy)$:
\BB
  \langle {\bf g} \rangle _{\bf x}(\bfx-\bfy)={1\over\overline
 {\cV}_{\bf x}}\int_{\cV_{\rm \bf X}}\langle{\bf g}
\rangle(\bfz -\bfy)d\bfz,\quad \bfx\in {\Omega} _i.
\EE
Then Eq. (3.12) is reduced to
 \BBEQ
\bfsi ({\bf x})=
\langle \bfsi\rangle ({\bf x})+\int_{w}\big[
\lle\!\lle {\bf \Gamma} (\bfx-\bfy) \bfeta\rle\!\rle_{\bf x}({\bf y})+
\lle\!\lle \bfL^c\nabla\bfG (\bfx-\bfy)
\bff_1\rle\!\rle_{\bf x}(\bfy)\big]d\bfy
+\langle\!\langle\bfcI^{\Gamma}_{\sigma}\rangle\!\rangle_{\bf x},
\EEEQ
where
\BB
\langle\!\langle\bfcI^{\Gamma}_{\sigma}
\rangle\!\rangle_{\bf x}=
\int_{\Gamma}\langle\!\langle{\bfGa}^{\Gamma}
{\bf (x-s)} \bfet^1 ({\bf s})\rle\!\rle_{\bf x}({\bf s})
[\varphi (v_i,{\bf x}_i
\vert v_1,{\bf x}_1, \ldots,v_n,{\bf x}_n)
-\varphi (v_i,{\bf x}_i)]d\bfs.
\EE
is a centered  surface integral and one introduces a new centering operation
$\lle\!\lle(\cdot)\rle\!\rle_{\bf x}$ such as, e.g.,
\BB
\lle\!\lle\bfGa(\bfx-\bfy)\bfet\rle\!\rle_{\bf x}(\bfy)=\bfGa(\bfx-\bfy)\bfet(\bfy)-
\lle\bfGa(\bfx-\bfy)\bfet\rle_{\bf x}(\bfy). 
\EE
 For the analysis of integral convergence in Eq. (4.7) at $|\bfx-\bfy|\to\infty$, we
assume that $\bfet^1 ({\bf y})$ can be regarded as a constant, equal to the value at $\bfx_i=\bfy_i$, and may thus be taken outside the averaging operation $\lle\!\lle(\cdot)\rle\!\rle_{\bf x}$. Then we expand $\bfGa(\bfz-\bfy)$  ($\bfz\in \cV_{\bfx}$) in a Taylor series about $\bfx$
and integrate term by term over the cell  $\cV_{\bf x}$ with the center $\bfx$
\BBEQ
\!\!\!\bfGa(\bfz-\bfy)\!\!\!&=&\!\!\!
\bfGa(\bfx-\bfy)+(\bfz-\bfx)\nabla\bfGa
(\bfx-\bfy)
+{1\over2}(\bfz-\bfx)\otimes(\bfz-\bfx)
\nabla\nabla\bfGa(\bfx-\bfy)
\ldots,\\
\!\!\!\!\!\!\!\!\!\!\!\langle\bfGa \rangle _{\bf x}(\bfx-\bfy)
\!\!\!\!\!&=&\!\!\!\!\!\bfGa(\bfx\!-\!\bfy)
+{1\over 2\overline {\cV}_{\bf x}}
\!\!\int_{\cV_{\bf x}}\!\!(\bfz\!-\!\bfx)\!\otimes\!(\bfz\!-\!\bfx)d\bfz
\nabla\nabla\bfGa(\bfx\!-\!\bfy)
\ldots.
\EEEQ
The similar expansion can be performed for the tensor
$\nabla\bfG(\bfx-\bfy)$ that leads to
\BBEQ
\!\!\! \!\!\!\langle\! \langle\bfGa\rangle\!\rangle  _{\bf x}(\bfx-\bfy)\!\!\!&=&\!\!\!
-{1\over 2\overline {\cV}_{\bf x}}
\!\!\int_{\cV_{\rm \bf x}}\!\!(\bfz\!-\!\bfx)
\!\otimes\!(\bfz\!-\!\bfx)d\bfz
\nabla\nabla\bfGa(\bfx-\bfy)+
\ldots .\\
\!\!\!\!\!\!\!\!\!\langle\! \langle\nabla{\bf G} \rangle\!\rangle  _{\bf x}(\bfx\!-\!\bfy)\!\!\!&=&\!\!\!
-{1\over 2\overline {\cV}_{\bf x}}
\!\!\int_{\cV_{\rm \bf x}}\!\!(\bfz\!-\!\bfx)
\!\otimes\!(\bfz\!-\!\bfx)d\bfz
\nabla\nabla\nabla{\bf G}(\bfx\!-\!\bfy)\!+
\ldots .
\EEEQ
As is evident from Eq. (4.12), the tensor
$\langle\! \langle{\bfGa} \rangle\!\rangle  _{\bf x}(\bfx-\bfy)$
is  of order  $O(a^2_{\cV}|\bfx-\bfy|^{-d-2})$
with the dropped terms in Eq. (4.12) being of order
 $O(a_{\cV}^4|\bfx-\bfy|^{-d-4})$ and higher order terms.
Then the absolute convergence of volume integral (4.7)
is assured because at sufficient distance $\bfx$ from the
boundary $\Gamma$ as $|\bfx-\bfy|\to\infty$ the integration over
$\bfy$ can be carried out
independently  for both $\langle\! \langle{\bf U} \rangle\!\rangle  _{\bf x}
(\bfx-\bfy)$  (the function of the ``slow" variable $\bfx-\bfy)$
 and  $\bfet(\bfy)$ (the function of ``fast" variable $\bfy$), and
therefore the volume integral converges absolutely.
In a similar manner the term
$\langle\!\langle{\bfGa}^{\Gamma}
{\bf (x-s)} \bfet^1 ({\bf s})\rle\!\rle_{\bf x}({\bf s})$  in the surface integral (4.8) is of order
$O(a_{\cV}^2|\bfx-\bfy|^{-d-1})$,
and the surface integral vanishes at $|\bfx
-{\bf s}|\to\infty$ , ${\bf s}\in \Gamma$ if
$\langle \bfsi\rangle(\bfs)$ and, therefore
$\langle \bfet^1\rangle_j (\bfs)$  grows
with $\bfs$ slower then $O(|\bfx-\bfs|^{2-\beta})$
($\beta={\rm const.}>0$).
For the same reason the volume integral with the integrand
$\langle\!\langle\bfL^c\nabla {\bf G}(\bfx-\bfy)\bff_1\rangle\! \rangle _{\bf x}
(\bfy)$ converges absolutely for bounded functions $\bff_1(\bfy)$.

By this means the locality principle exists in
the new Eq. (4.7) for the case of long-range order composites
being considered if the average stress $\langle \bfsi\rangle(\bfs)$
grows with $\bfs$ slower then $O(|\bfx-\bfs|^{2-\beta})$
($\beta={\rm const.}>0$).

\bigskip
\noindent {\fourbf 5. Conclusion}
\medskip

As noted in Introduction, the final goals of micromechanical research of composites
involved in  a prediction of both the overall effective properties
 and statistical moments of stress-strain fields are based on the
approximate solution of the exact initial  integral equation (2.8).
Absolute convergence of integrals in Eq. (2.8) is provided by different versions of the centering procedure performed for the different cases of the boundary conditions, microtopology of composite material, and accompanied assumptions. It leads to the different general Eqs. (3.2), (3.6), and (3.17) (and their stress-state analogs) called the initial  integral equations which have the different renormalizing terms. Then, considering some conditional ensemble averages of the general equations either (3.2), (3.6), or (3.17) yield the infinite hierarchy of equations. These truncated hierarchies of equations are solved as a system of coupled equations. One starts with the last members of these hierarchies, the ones which have the most inclusions held fixed, because these equations does not depend on the others. The fields so obtained give the previous terms in the next equations up the hierarchies. One continues step by step up the hierarchies until the unconditionally averaged fields are finally obtained. However, these standard procedures (differing by both the numbers of coupled equations and assumptions exploited for their solutions) have  fundamentally diverse backgrounds defined by the features of the renormalizing terms in Eqs. (3.2), (3.6), and (3.17).

These features of the initial integral equations are fundamental for subsequent solving the truncated hierarchy involving a rearrangement of each appropriate equation before it is solved. The most successful rearrangement are those which make the right-hand side of the coupled equations reflect the detailed corrections to that basic physics.
So, the advantage of Eq. (3.2) with respect to Eq. (3.30) (even for the case when Eq. (3.30) is correct) is explained by faster convergence of corresponding integrals as $|\bfx-\bfy|\to \infty$. The centering method realized at the obtaining of Eq. (3.2) subtracts the difficult state at infinity from the equation, i.e. roughly speaking the constant force-dipole density  expressed through an alternative technique of the Green's function. This dictates the fundamental limitation of possible generalization of Eq. (3.2) to both the FGMs and inhomogeneous boundary conditions. The mentioned deficiency of Eq. (3.2) was resolved by Eq. (3.6) which renormalizing term provides an absolute convergence of the integral in Eq. (3.6) at $|\bfx-\bfy|\to \infty$ for general cases of the FGMs. However, the same term in Eq. (3.6) is used in a short-range domain $|\bfx-\bfy|<3a$ in the vicinity of the point $\bfx\in w$. A fundamental deficiency of Eq. (3.6) is a dependence of the renormalizing term
$\bfU(\bfx-\bfy)\lle\bfL_1\bfep\rle(\bfy)$ [obtained in the framework of the
asymptotic approximation (3.23)] only on the statistical average $\lle\bfL_1\bfep\rle(\bfy)$ rather than $\lle\bfU(\bfx-\bfy)\bfL_1\bfep\rle(\bfy)$ in Eq. (3.17).
What seems to be only a formal trick [abandoning the use of the approximations of the kind
(3.23) and (3.24)] is in reality a new background of micromechanics
(3.17)-(3.19)
[which does not use an approximation of the kind either (3.23), (3.24), or (3.3) as in Eqs. (3.2) and (3.6)]
yielding the discovery of fundamentally new effects even in the theory of
statistically homogeneous media subjected to homogeneous boundary conditions (see for details the accompanied paper by Buryachenko [48]). 


\medskip
\noindent{\bf Acknowledgments:}
\medskip

This work was partially supported by both the Visiting Professor Program of  the University of Cagliari funded by Region Autonoma della Sardegna and the  Eppley Foundation for Research.

\medskip
\noindent{\bf References}
\medskip
{\baselineskip=9pt
\parskip=1pt
\lrm

   \hangindent=0.4cm\hangafter=1\noindent
1. {Plankensteiner, A.F.}, {B\"ohm, H.J.},
{Rammerstorfer F.G.},
 and {Buryachenko,  V.A}.:
 Hierarchical modeling of the mechanical behavior
 of high speed steels as layer--structured particulate MMCs.
{\tenitr Journal de Physique IV}, {\tenbf 6}, C6-395--C6-402 (1996)

\hangindent=0.4cm\hangafter=1\noindent
2. {Plankensteiner, A.F., B\"ohm, H.J.,
   Rammerstorfer, F.G., Buryachenko, V.A., Hackl,G}.:
Modeling of layer--structured high speed tool steel.
{\tenitr Acta Metall. et Mater.}, {\tenbf 45}, 1875--1887 (1997)

\hangindent=0.4cm\hangafter=1\noindent
3. {Conlon, K.T.} and {Wilkinson, D.S.}:
Microstructural inhomogeneity and the strength of particulate metal
matrix composites. {\tenitr IUTAM Symp. on Micromechanics of
Plasticity and Damage of Multiphase Materials}, pp. 347--354, eds.,
 A. Pineau \& A. Zaoui, Kluver Academic Publ., Dordrecht (1967)

\hangindent=0.4cm\hangafter=1\noindent
4. Krajcinovic, D.: {\tenitr Damage Mechanics}. Elsevier, Amsterdam (1996)

 \hangindent=0.4cm\hangafter=1\noindent
5. {Markworth, A.J.}, {Ramesh, K.S.}, {Parks, W.P.}
          Review. Modeling studies applied to functionally graded materials. {\tenitr J. Materials Science}, {\tenbf 30}, 2183--2193 (1995)

\hangindent=0.4cm\hangafter=1\noindent
6. {Mortensen, A.}, {Suresh S.}
 Functionally graded metals and
metal--ceramic composites: Part 1. Processing. {\tenitr Int. Mater. Reviews,} {\tenbf 40}, 239--265 (1995)

\hangindent=0.4cm\hangafter=1\noindent
7. Erdogan, F.:  Fracture mechanics of functionally graded materials.
{\tenitr Compos. Engng.}, {\tenbf 5}, 753--770 (1995)

\hangindent=0.4cm\hangafter=1\noindent
8. Praveen, G. N., Reddy, J. N.: Nonlinear transient thermoelastic analysis of functionally
graded ceramics-metal plates. {\tenitr
Int. J. Solids Structures}, {\tenbf 35}, 4437--4476 (1998)

\hangindent=0.4cm\hangafter=1\noindent
9. {Buryachenko, V.A.}, {Rammerstorfer, F.G.}:
Micromechanics and nonlocal effects
in graded random structure matrix composites.
{\tenitr IUTAM Symp. on
    Transformation Problems in Composite and Active Materials},
pp. 197--206, eds.  Y.A. Bahei-El-Din \& G. J. Dvorak, Kluwer
Academic Publ., Dordrecht (1998)

\hangindent=0.4cm\hangafter=1\noindent
10. Buryachenko, V.A.:{\tenitr Micromechanics of Heterogeneous Materials}. Springer, NY
 (2007)

\hangindent=0.4cm\hangafter=1\noindent
11. Khoroshun, L.P.: Random functions theory in problems on the macroscopic
characteristics of microinhomogeneous media. {\tenitr Priklad Mekh}, {\tenbf 14}(2),3-–17
(In Russian. Engl Transl. {\tenitr Soviet Appl Mech}, {\tenbf 14}, 113-–124) (1978)

\hangindent=0.4cm\hangafter=1\noindent
12. {Shermergor, T.D.}:  { \tenitr The Theory of Elasticity  of Microinhomogeneous Media}. Nauka, Moscow (1977)  (In Russian).

\hangindent=0.4cm\hangafter=1\noindent
13. {Willis, J.R.}:
Variational and related methods for the overall properties
of composites. {\tenitr Advances in Applied Mechanics},
 {\tenbf 21}, 1--78 (1981)

\hangindent=0.4cm\hangafter=1\noindent
14. Lekhnitskii, A.G.:  {\tenitr Theory of Elasticity of an Anisotropic Elastic Body}. Holder Day, San Francisco (1963)

 \hangindent=0.4cm\hangafter=1\noindent
 15. Luciano, R., Willis, J.R.:  Boundary-layer correlations for stress and strain field
  in randomly heterogeneous materials. {\tenitr J Mech Phys Solids}, {\tenbf 51}, 1075-–1088
(2003)

 \hangindent=0.4cm\hangafter=1\noindent
16.  Luciano, R., Willis, J.R.:  Non-local constitutive equations for functionally
  graded materials. {\tenitr Mech Mater}, {\tenbf 36}, 1195-–1206 (2004)

\hangindent=0.4cm\hangafter=1\noindent
17. Torquato, S.:  {\tenitr Random Heterogeneous Materials: Microstructure and Macroscopic Properties}. Springer-Verlag, New York, Berlin (2002)

\hangindent=0.4cm\hangafter=1\noindent
18. Quintanilla, J., Torquato, S.:  Microstructure functions for a model of statistically
inhomogeneous random media.
{\tenitr  Physical Review E}, {\tenbf 55}, 1558--1565  (1997)

\hangindent=0.4cm\hangafter=1\noindent
19. McCoy, J.J.:  Macroscopic response of continue with random microstructure.
In: Nemat-Nasser S (ed) {\tenitr Mechanics Today}. Pergamon Press, Oxford, {\tenbf 6}:1–40
(1981)

\hangindent=0.4cm\hangafter=1\noindent
20. Brebbia, C.A., Telles, J.C.F., Wrobel, L.C.:
{\tenitr Boundary Element Techniques}. Springer-Verlag, Berlin (1984)

\hangindent=0.4cm\hangafter=1\noindent
21. {Gel`fand, I.A.}, {Shilov, G.}:
{\tenitr Generalized Functions.  Vol. I}. Academic Press, NY (1964)

 \hangindent=0.4cm\hangafter=1\noindent
 22. Lipinski, P., Berveiller, M., Reubrez, E., Morreale, J.:
 Transition theories of elastic-plastic deformation of
 metallic polycrystals. {\tenitr Archive of Appl. Mechanics},  {\tenbf 65}, 295-311
(1995)

 \hangindent=0.4cm\hangafter=1\noindent
23.  Ju, J. W., Tseng, K. H.:  A
 three-dimensional micromechanical theory for brittle solids
 with interacting microcracks. {\tenitr Int. J. Damage
 Mechanics} {\tenbf 1}, 102--131 (1992)

 \hangindent=0.4cm\hangafter=1\noindent
24. Ju, J. W., Tseng, K. H.:
 Improved two-dimensional micromechanical theory for brittle solids with
 randomly located interacting microcracks.
 {\tenitr Int. J. Damage Mechanics} {\tenbf 4}, 23--57 (1995)

 \hangindent=0.4cm\hangafter=1\noindent
25. Fassi--Fehri, O., Hihi, A., Berveiller, M.:
 {Multiple site self consistent scheme.}
 {\tenitr Int. J. Engng. Sci}. {\tenbf 27}, 495--502 (1989)

\hangindent=0.4cm\hangafter=1\noindent
26. Batchelor, G.K.: Sedimentation in a dilute dispersion of spheres.
{\tenitr J Fluid Mech}, {\tenbf 52}, 245-–268 (1972)

\hangindent=0.4cm\hangafter=1\noindent
27. Jeffrey, D.J.:  Conduction through a random suspension of spheres.
{\tenitr Proc Roy Soc Lond}, {\tenbf A335}, 355-–367 (1973)

\hangindent=0.4cm \hangafter=1\noindent
28.  Chen, H.S., Acrivos, A.:
The effective elastic moduli of composite materials containing spherical
inclusions at non-dilute concentrations.
{\tenitr Int. J. Solids Structures}, {\tenbf 14}, 349-364 (1978)

\hangindent=0.4cm\hangafter=1\noindent
29. Willis, J.R., Acton, J.R.:  The overall elastic moduli of a dilute suspension of
spheres. {\tenitr Q J Mechan Appl Math}, {\tenbf 29}, 163–-177 (1976)

\hangindent=0.4cm\hangafter=1\noindent
30. Zeller, R., Dederichs, P.H.:  Elastic constants of polycrystals.
{\tenitr Phys. Stat. Sol.}, {\tenbf a55}, 831--842 (1973)

\hangindent=0.4cm\hangafter=1\noindent
31.  O'Brian, R.W.:  A method for the calculation of the effective
transport properties of suspensions of interacting particles. {\tenitr J.
Fluid. Mech.} {\tenbf 91}, 17--39 (1979)

\hangindent=0.4cm\hangafter=1\noindent
32. McCoy, J.J.:  On the calculation of bulk properties of heterogeneous materials. {\tenitr Quarterly of Applied Math.} {\tenbf 36}, 137--149 (1979)

\hangindent=0.4cm\hangafter=1\noindent
33.  Kr\"oner, E.:   On the physics and mathematics of self-stresses. In: Zeman, J.L. and Ziegler, F. {\tenitr Topics in Applied Continuum Mechanics}. Springer-Verlag, Wien, 22--38
(1974)

\hangindent=0.4cm\hangafter=1\noindent
34.  Kr\"oner, E.:   Statistical modeling.
    In: Gittus~J,  Zarka~J (eds),    {\tenitr   Modeling Small Deformations of Polycrystals.}
  Elsevier, London/NY,  229--291 (1986)

\hangindent=0.4cm\hangafter=1\noindent
35.  Kr\"oner, E.: Modified Green function in the theory of
heterogeneous and/or anisotropic linearly elastic media. {\tenitr
Micromechanics and Inhomogeneity. The Toshio Mura 65th Anniversary Volume}, pp. 197--211,
eds., G. J. Weng, M. Taya, H. Abe, Springer--Verlag, NY  (1990)

\hangindent=0.4cm\hangafter=1\noindent
36. {Kanaun, S.K.}:  Self-consistent field approximation for an elastic composite medium.
{\tenitr Zhurnal Prikladnoi i Tehknich. Fiziki}, {\tenbf 18}, (2), 160--169 (1977)
(In Russian. Engl. Transl. {\tenitr J. Applied Mech. Techn. Physics}, {\tenbf 18}, 274--282.\
(1977))

\hangindent=0.4cm\hangafter=1\noindent
37. Kanaun, K.K., Levin, V.M.:  {\tenitr Self-Consistent Methods for Composites}.
Vol. 1, 2. Springer, Dordrecht (2008)

\hangindent=0.4cm\hangafter=1\noindent
38. Buryachenko, V.A.:  Locality principle and general integral equations of
micromechanics of composite materials. {\tenitr Math Mech Solids}, {\tenbf 6}, 299-–321
(2001)

\hangindent=0.4cm\hangafter=1\noindent
39. Buryachenko, V.A.,  Parton, V.Z.: Effective  parameters
of statistically inhomogeneous matrix composites.
{\tenitr  Izv. AN SSSR, Mekh. Tverd. Tela.} (6), 24--29  (1990). (In Russian.
Engl. Transl. {\tenitr  Mech. Solids} {\tenbf  25}, 22--28  (1990))

 \hangindent=0.4cm\hangafter=1\noindent
40. Buryachenko,~V.A.:    Some nonlocal effects in graded random structure matrix
composites.  {\tenitr     Mech Res Commun,} {\tenbf      25},117--122 (1998)

\hangindent=0.4cm\hangafter=1\noindent
41. {Filatov, A.N.}, {Sharov, L.V.}: (1979)
{\tenitr Integral Inequalities and the Theory of Nonlinear Oscillations.}
 Nauka, Moscow (1979) (In Russian).

\hangindent=0.4cm\hangafter=1\noindent
42. {Willis, J.R.}:
Variational principles and bounds for the overall properties of
composites. {\tenitr Continuum Models of disordered systems},
pp. 185--215, ed., J. W. Provan, University of Waterloo Press, Waterloo (1978)

\hangindent=0.4cm\hangafter=1\noindent
43. Hansen, J.P., McDonald, I.R.:  {\tenitr Theory of Simple Liquids.} Academic Press, NY
(1986)

\hangindent=0.4cm\hangafter=1\noindent
44. Torquato, S., Lado, F.: Improved bounds on the effective elastic moduli of
cylinders. {\tenitr J. Applied Mechanics} {\tenbf 59}, 1--6 (1992)

\hangindent=0.4cm\hangafter=1\noindent
45. Lipinski, P.,  Berveiller, M.:
 Elastoplasticity of micro-inhomogeneous metals at large strains.
 {\tenitr Int. J. Plasticity}, {\tenbf 5}, 149--172  (1989)

\hangindent=0.4cm\hangafter=1\noindent
46. Eshelby, J.D.: The determination of the elastic field of an ellipsoidal inclusion,
and related problems. {\tenitr Proc Roy Soc Lond}, {\tenbf A241}, 376-–396  (1957)

\hangindent=0.4cm\hangafter=1\noindent
47. Beran, M.J., McCoy, J.J.: Mean field variations in a statistical sample of
heterogeneous linearly elastic solids. {\tenitr Int J Solid Struct}, {\tenbf 6}, 1035-–1054
(1970)

\hangindent=0.4cm\hangafter=1\noindent
48. Buryachenko, V.A.:
On the thermo-elastostatics of heterogeneous materials.
II. Analyze and generalization of some basic hypotheses and propositions.
{\tenitr Acta Mech}. (2009)
(Submitted).

 \hangindent=0.4cm\hangafter=1\noindent
 49. Drugan, W. J., Willis, J. R.:
   A micromechanics-based nonlocal constitutive equation
  and estimates of representative volume elements for elastic composites.
  {\tenitr J Mech Phys Solids}, {\tenbf 44}, 497-–524 (1996)

 \hangindent=0.4cm\hangafter=1\noindent
50.  Drugan, W. J.:  Micromechanics-based variational estimates for a higher-order
  nonlocal constitutive equation and optimal choice of effective moduli for elastic
  composites. {\tenitr J Mech Phys Solids}, {\tenbf 48}, 1359--1387 (2000)

 \hangindent=0.4cm\hangafter=1\noindent
51. Drugan, W. J.:  Two exact micromechanics-based nonlocal constitutive equations  for random linear  elastic composite materials. {\tenit J Mech Phys Solids}, {\tenbf  48}, 1359--1387 {\tenitr J Mech Phys Solids}, {\tenbf 51}, 1745--1772 (2003)

 \hangindent=0.4cm\hangafter=1\noindent
52.  Sharif-Khodaei, Z, Zeman, J.:  Microstructure-based modeling of elastic
 functionally graded materials:  one dimensional case. {\tenitr J. Mechanics of
   Materials and Structures}, {\tenbf 3}, 1773--1796 (2008)

\hangindent=0.4cm\hangafter=1\noindent
53. Torquato, S.: Effective stiffness tensor of composite media -- I.
Exact series expansion. {\tenitr J. Mech. Phys. Solids},
{\tenbf 45}, 1421--1448 (1997)

\hangindent=0.4cm\hangafter=1\noindent
54. Xu, X.F.: Generalized variational principles   for uncertainty quantification of boundary value problems of random   heterogeneous materials. {\tenit J. Eng. Mech.}, {\tenbf 135}, 1180-1188 (2009)

}
\end{document}